# Patterns to analyze requirements of a Decisional Information System

SABRI AZIZA

ENSIAS, University Mohammed-V-Souissi, Rue Mohammed Ben Abdellah Regragui, B.P. 713

Agdal, Madinat Al Irfane

Rabat, Morocco

E-mail : azizasabri@yahoo.com

KJIRI LAILA

ENSIAS, University Mohammed-V-Souissi, Rue Mohammed Ben Abdellah Regragui, B.P. 713

Agdal, Madinat Al Irfane

Rabat, Morocco

E-mail : kjiri@ensias.ma

## ABSTRACT

The domain of analysis and conception of Decisional Information System (DIS) is, highly, applying new techniques and methods to succeed the process of the decision and minimizing the time of conception. Our objective in this paper is to define a group of patterns to ensure a systematic reuse of our approach to analyse a DIS's business requirements. We seek, through this work, to guide the discovery of an organization's business requirements, expressed as goals by introducing the notion of context, to promote good processes design for a DIS, to capitalize the process and models proposed in our approach and systematize reuse steps of this approach to analyze similar projects or adapt them as needed. The patterns are at the same time the process's patterns and product's patterns as they capitalize models and their associated processes. These patterns are represented according to the P-SIGMA formalism.

## 1. INTRODUCTION

The success of a business intelligence project is strongly associated to the step of analysis requirements, situated upstream of the development process which consists of defining the project environment and all data and processing which are required in process of development. Despite the impact of this phase throughout the development process, many problems related to the incoherence, the semantic ambiguity and the difficulty of modeling requirements were detected. So, we have solved those problems by proposing an analysis approach that guide analysts throughout the process of DIS's designing [2]. This approach incorporates the concept of business context of the organization which consists of defining organization structure as well as its actors and to identify the business requirements of each actor in a predefined context.

Moreover, in a business intelligence project, it is often necessary to reuse the tasks of the process of requirements analysis empirically by the DIS's designers. Thus, the extraction of clear criteria for reuse is a hard task that depends on how to specify and identify suitable components for reuse. In this sense, we opt for an approach guided by the patterns for the analysis of business requirements of a DIS, that consists in proposing a set of process and product patterns reusable and flexible product, which we associate a reuse strategy to ensure adaptation. The reuse of these patterns will be more systematic, since it capitalizes the processes and models.

The paper is organized the following way. First, we present a state of the art approaches for analyzing business requirements of a DIS. In Section 3, we briefly present approaches based patterns. In Section 4, we propose a basic pattern for the analysis of DIS's business requirements. Finally, we end our article by a conclusion and perspectives of our work.

## 1. APPROACHES FOR BUSINESS REQUIREMENTS ANALYSIS OF A DIS

### 2.1 The phase of requirements analysis of a DIS

The step of requirements analysis is the key to any successful software project that significantly reduces the risk of project failure. Requirements specification of a DIS data warehouse project determines the different functions of the data warehouse and all the required information that it should cover [12]. The modeling approach of DIS's requirements (collected from source and / or users) is based on the following steps [5]: the gathering collect requirements, analysis requirements, validate and model requirements.

In this section we will define reference criteria, in order to compare approaches of requirements analysis of a DIS, and to situate our work relatively to these approaches.

### 2.2 Criteria of evaluation

We chose to evaluate approaches for analyzing or expressing requirements and for designing and manipulating design's reusable components. This reference criterion proposes a set of criteria to measure the similarities and differences between these works, by detecting their weak points. We cite all the suggested criteria: Completeness of analysis approach, Means of requirements gathering, Models of requirements formulation, Coherence and semantics of requirements, Orientation and support for users on their expression's requirements, Distinction of the actors of DIS, Classification of collected requirements, Treatment of requirements, Capitalization and reuse of knowledge, Documentation of process analysis of the decisional requirements.

After defining the criteria enabling us to compare approaches to requirements analysis of a DIS, we will present a state of the art practices.

### 2.3 State of the art

In the literature, three groups of models are proposed for the representation of a DIS requirements including: existing models (entity-relationship, if used) [4], requests to represent the requirements [9] and models of goals [11][8] in where authors represent the requirements according to the goals they seek. Among the models used for the formalization of requirements, none is similar to the multidimensional representation of data made by the decision makers [6] Moreover, these models don't clarify in a systematic way the extraction of business intelligence data (facts and dimensions) and don't take into account context to collect of an organization's business requirements.

Thus, the works based on existing models are not familiar to decision makers and they don't distinguish between strategic decision-makers and tactical decision-makers. As for approaches based on requests, structuring the form of request is not familiar to common user makers who hardly validate the specification of their requirements. Goals based works don't distinguish between different types of decision makers and use models formalizing requirements which are not close to the multidimensional representation of data expressed by decision makers. Therefore, some difficulties remain to validate the formalization of requirements before designing the schema of the DIS.

The results of comparison of analysis works for DIS according to the criteria that we have defined above are illustrated in the following table:

**Table 1: Comparative approaches to requirements analysis of a DIS**

| Approaches of requirements analysis of SID | Mazon and al., 2005 | Soussi and al., 2005 | Ghozzi and al., 2005 | Annoni, 2007 | El Golli, 2008 |
|---|---|---|---|---|---|
| Completeness of analysis approach | Tasks | Tasks |  | Tasks | Overall |
| Means of requirements gathering | Natural language | Natural language | Natural language | Table | Natural language |
| Models of requirements formulation | I* (model of goal) | UML (Diagram of classe) | Request | Table | MAP (but) |
| Coherence and semantics of requirements | Untreated | Untreated | Untreated | Untreated | Untreated |
| Orientation and support for users on their expression's requirements | Explicit but incomplete | Explicit but incomplete | Explicit but incomplete | Explicit but incomplete | Explicit but incomplete |
| Distinction of the actors of SID | yes |  |  | yes | yes |
| Classification of collected requirements |  |  |  | yes | yes |
| Treatment of requirements | Implicite |  |  | Implicit | Explicit but incomplete |
| Capitalization and reuse of knowledge |  | yes (diagram classe) |  | yes (patterns) |  |
| Documentation |  |  |  | yes |  |

We conclude that these approaches are limited. Thus, they don't treat the coherence and semantic of a DIS's requirements and they don't take into account the structure of the organization or the necessity to define its activities. Moreover, no approach had proposed the encapsulation of requirements in a predefined context and has explained the automatic extraction of facts and dimensions. Also, the concept of reuse is obscured in most of these approaches, except [6] who proposed a catalog of patterns as a product patterns and process patterns. In the following text, we will present a state of the art approaches based on patterns.

## 2. APPROACHES BASED ON PATTERNS

In our context, a pattern is defined as "a fully realized form, original or model accepted or proposed for imitation; something that is seen as a normative example can be copied, used as an example or archetype" [17].

### 3.1 Classification of patterns

In the literature, we identify three principal types of patterns: analysis patterns, design patterns and implementation patterns. Analysis pattern identifies recurrent problems in the expression of requirements of different application domains. The patterns' analysis [16] is an example of patterns used in the phase of requirement analysis. Design patterns [7] identify, appoint and abstract out common themes in the area of object-oriented design. Implementation patterns are generally specific to a programming language by describing the implementation of particular aspects of components or the relations between them in a particular programming language [13].

### 3.2 The formalism pattern P-SIGMA

P-Sigma formalism [1] is an attempt to unify structured formalisms that have been proposed [17][7]. We have chosen this formalism as the basis of our representation because it incorporates the products aspects of the expression and processes as well a large number of inter-patterns. Thus, we present the P-Sigma formalism chosen to represent our approach of the requirements analysis in decision-making [2]. Our goal through this formalism is to facilitate the selection, organization and reuse patterns. They allow the definition of many information related to the implementation of the solution-approach this consists on the realization of activity diagrams and patterns.

## 4. THE PRODUCT OF THE BUSINESS REQUIREMENTS ANALYSIS PHASE OF DIS: PROPOSITION

### 4.1 The pattern « Diagnosis of the organization »

The diagnosis of the organization is consists of three tasks: (i) define the business of the organization, (ii) determine the activities defining the job and (iii) list contexts associated with each activity. Thus, we propose the following activity diagram which illustrates the process of defining the business of the organization:

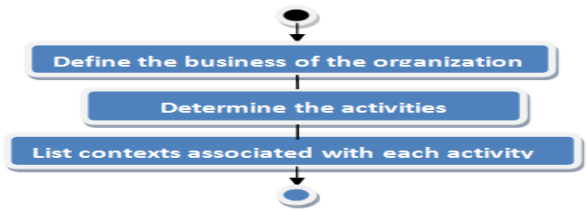

**Fig 1: Activity diagram "Defining the organization business"**

After defining the business context of the organization [2] we have the step of collecting of requirements. Thus, we propose the following model for the systematic collection of information concerning the organization. This is the model "Diagnosis of the Organization":

| Business of the organization |
|---|
| Activity 1 |
| List 1of contexts: context1, context2, context3… |
| Activity 2 |
| List 2 of contexts: context1, context2, context3… |
| … |
| Activity n : |
| List n of contexts: context1, context2, context3… |

**Fig 2: Model «Diagnosis of the Organization»**

This model serves as a document of diagnosis for the organization. It allows to identify all activities related to the business of the organization and the contexts associated with each activity.

To ensure capitalization and reuse of this model, we establish the pattern "Diagnosis of the organization" as follows:

**Table 2: The pattern « Diagnosis of the Organization»**

| Part | Rubric | Fields |
|---|---|---|
| Interface | Symbol | DiagnosisSID |
| | Name | Diagnosis of business organization |
| | Classification | DIS ^Analysis^Product^Process |
| | Context | This pattern is reused in the definition of business organization. |
| | Problem | Guide the discovery of an organization's activities and contexts associated with them. |
| | Strength | This pattern details the steps to determine the contexts associated with the activities of the organization. |
| Réalization | Process Solution | The process solution consists in the achievement of activity diagram "Defining the business organization" as follows:<br><br>Define the business of the organization → Determine the activities → List contexts associated with each activity |
| | Model Solution | The model solution obtained during the application of activity diagram is the model "Diagnosis of the Organization" as follows:<br><br>Business of the organization / Activity 1 / List 1of contexts: context1, context2, context3… / Activity 2 / List 2 of contexts: context1, context2, context3… / … / Activity n : / List n of contexts: context1, context2, context3… |
| Relationship | Uses | Process pattern "Analyzing the requirements of a DIS " [2] |

This pattern encapsulates the activity diagram "Defining the business of the organization" that guides the task of defining activities and contexts of the organization, also the model "Diagnosing of the organization" which allows their representation.

## 3.1 The pattern "Identification of business requirements"

The collection aims to understand the area that must be modeled. In our approach to requirements assessment decision [2], we have adopted a use case diagram to determine the actors of DIS and their goals. The actors of the diagram are actors who predefined in the DIS (strategic, tactical and system) and the use cases represent the goals defined by each actor.

In addition, the task of collecting business requirements is to achieve three main tasks: (i) define the actors of the organization, (ii) identify the requirements of each actor, and (iii) establish the use cases "Expression's requirements of DIS". Thus we present the following activity diagram which illustrates the process of identifying business requirements:

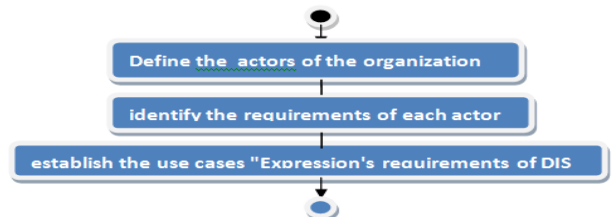

**Fig 3: Activity Diagram « Identifying a DIS's business requirements »**

To ensure capitalization and reuse of this model, we establish the pattern process "Identifying business requirements" as follows:

**Table 3: The pattern « Identification of business requirements »**

| Part | Rubric | Fields |
|---|---|---|
| Interface | Symbol | RecenseBesoinsSID |
| | Name | Identification of business requirements |
| | Classification | SID ^ Analysis ^ Process ^ Product |
| | Context | This pattern is reused in a new collection of a DIS's business requirements. |
| | Problem | To guide the collection of a DIS's business requirements. |
| | Strength | This pattern describes how to identify a DIS's business requirements. |
| Realization | Process Solution | It consists in the achievement of the activity diagram "Identifying a DIS's business requirements" as follows: 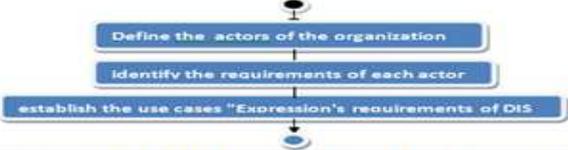 |
| | Model Solution | The model solution obtained when applying the process is the following use case model " Expression's requirements of DIS": 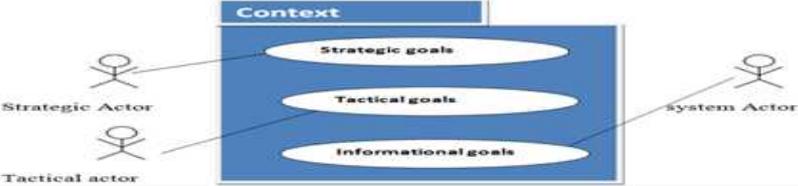 |
| Relationship | Uses | |
| | Requires | The pattern «Diagnosis of the Organization» |

This pattern capitalizes both the solution approach which consists in the realization of the activity diagram "Identifying business requirements" and the model solution "Expression's requirements of a DIS" obtained who applying the approach.

## 4.3 Pattern "Association of the strategic goals to the tactical goals»

In our approach [3], business requirements are collected in the form of goals. This collecting is based in principle on a formulation of goals in a natural language. The classification of decision-making requirements, identified as goals, consists of organizing them into three types (strategic, tactical and informational). This classification will allow us to make explicit the links between goals in a given context. After this classification of goals, we proceed to the association of each strategic goal to all tactical goals attached to it according to a predefined context. This association is carried out according to the following model:

| Activity | | | |
|---|---|---|---|
| context | | | |
| Strategic goal 1 | Strategic goal 2 | ... | Strategic goal n |
| List 1 of tactical goals | List 2 of tactical goals | ... | List n of tactical goals |

**Fig 4: Association model of strategic goals to tactical goals**

For the reasons of capitalization and to ensure possible reuse, we propose the following pattern:

**Table 4: The pattern «Association of the strategic goals to the tactical goals»**

| Part | Rubric | Fields |
|---|---|---|
| Interface | Symbol | AssocieButStra-ButTact |
| | Name | Association of the strategic goals to the tactical goals |
| | Classification | SID ^ Analysis ^ Product |
| | Context | This pattern is reused in the Association of the strategic goals to tactical goals. |
| | Problem | To guide the treatment of business goals and associated strategic goals to tactical goals. |
| | Strength | This pattern describes how to assemble the tactical goals associated with each strategic goal. |
| Réalization | Process Solution | From the use case "Expression's requirements of DIS" established the analyst can classify goals into three types (strategic, tactical and informational) and then associate each strategic goal to all tactical goals that are associated with him according to a predefined context. |
| | Model Solution | The model solution obtained from the application of process steps is the following use case model " Association model of strategic goals to tactical goals": 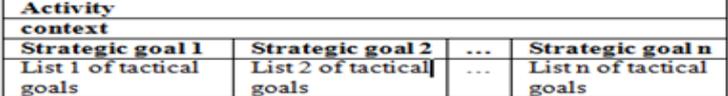 |
| Relationship | Uses | The pattern «Diagnosis of the Organization» |
| | Requires | The pattern «Identification of business requirements» |

The pattern "Association of the strategic goals to the tactical goals "capitalizes model" Model Association of strategic goals to the tactical goals ".

## 4.4 Pattern Association of the tactical goals for informational purposes

After establishment of the association models of strategic goals to the tactical goals, we proceed in the same way by associating each tactical goal with all informational purposes which are attached to it according to a predefined context. This association is performed according to the following table:

Fig 5: Association model of tactical goals to informational goals

| Activity | | | |
|---|---|---|---|
| Context | | | |
| Strategic goal | | | |
| Tactical goal 1 | Tactical goal 2 | ... | Tactical goal n |
| List 1 of informational goals | List 2 of informational goals | ... | List n of informational goals |

For the purpose of reuse, we capitalize the association model of the tactical goals to informational goals into the following pattern':

Table 5: The pattern «Association of the tactical goals to the informational goals»

| Part | Rubric | Fields |
|---|---|---|
| Interface | Symbol | AssocieButTact-ButInfo |
| | Name | Association of the tactical goals to the informational goals |
| | Classification | SID ^ Analysis ^ Product |
| | Context | This pattern is reused when associating the tactical goals to informational goals. |
| | Problem | To guide the treatment of business goals and the association of the tactical to informational goals. |
| | Strength | This pattern describes how to assemble informational goals associated with each tactical goal. |
| Realization | Process Solution | From the use case "Expression's requirements of DIS" established, the analyst can classify goals into three types (strategic, tactical and informational), then associate each strategic goal to all its tactical goals, then select a strategic goal and associated it with each of its tactical goals and select each tactical goal in order to associate it to all informational goals associated with it according to a predefined context. |
| | Model Solution | The model solution obtained during the application of activity diagram «Association model of tactical goals to informational goals»<br><br>| Activity | | | |<br>|---|---|---|---|<br>| Context | | | |<br>| Strategic goal | | | |<br>| Tactical goal 1 | Tactical goal 2 | ... | Tactical goal n |<br>| List 1 of informational goals | List 2 of informational goals | ... | List n of informational goals | |
| Relationship | Uses | The pattern : «Diagnosis of the Organization» |
| | Requires | The patterns : The pattern «Identification of business requirements» and «Association of the strategic goals to the tactical goals» |

The following section describes the formalization informational goals and then presents the pattern "Formalizing informational goals "to ensure the reuse.

## 4.5 Pattern "Formalization of informational goals"

To facilitate the task of the analyst of DIS to identifying business intelligence data, we defined a new version of the model informational goals [15]. We introduced the notion of paramètres_faits and the paramètres_dimensions [3]. Each informational goal is formulated with a "verb", a section called "Paramètres_faits" which contains the component of the fact table and a second section called "Paramètres_dimensions" which contains the parameters of dimension tables.

To make easy, the task of analysis, for analysts, we propose the following model formalization that allows to easily extracting business intelligence data:

| Activity | | |
|---|---|---|
| Context | | |
| Strategic goal | | |
| Tactical goal 1 | | |
| Verb | Fact_parameters | Dimension_parameters |
| V1 | FP1 | DP1 |
| V2 | FP2 | DP2 |
| V3 | FP3 | DP3 |

Fig 7: Model formalization of informational goals.

We capitalize the model and its settlement process into a pattern called "Formalization of informational goals":

Table 6: The pattern «Formalization of informational goals»

| Part | Rubric | fields |
|---|---|---|
| Interface | Symbol | FormaliseButInfo |
| | Name | Formalization of informational goals |
| | Classification | SID ^ Analysis ^ Product |
| | Context | This pattern is reused when analyzing informational goals.: extracting facts and dimensions. |
| | Problem | To guide the extraction of facts and dimensions after the formalization informational purposes. |
| | Strength | This pattern describes how to formalize informational goals associated with each tactical goal which is in turn associated with a strategic goal in a predefined context. |
| Realization | Process Solution | From the use case " Expression's requirements of DIS " established, the analyst can classify goals into three types (strategic, tactical and informational), then associate each strategic goal to all its tactical goals , then select a strategic goal and associated it with each of these tactical goals and select each tactical goal and associate it with all informational goals associated with it according to a predefined context, and finally formalize informational goals by Fact_parameters and Dimension_parameters. |
| | Model Solution | The model solution obtained during the application of the process is the following "Model formalization of informational goals":<br><br>| Activity | | |<br>|---|---|---|<br>| Context | | |<br>| Strategic goal | | |<br>| Tactical goal 1 | | |<br>| Verb | Fact_parameters | Dimension_parameters |<br>| V1 | FP1 | DP1 |<br>| V2 | FP2 | DP2 |<br>| V3 | FP3 | DP3 | |

| Relationship | Uses | The pattern : «Diagnosis of the Organization » |
|---|---|---|
| | Requires | The patterns : The pattern «Identification of business requirements», «Association of the strategic goals to the tactical goals» and «Association of the tactical goals to the informational goals» |

The pattern capitalizes "Model formalization of the informational goals" which clearly explains the facts and dimensions.

## 3. CONCLUSION AND PERSPECTIVES

Through the work presented in this paper, we proposed models accompanied by diagrams activity which guide their establishment. Initially, we presented an explicit analytical approach and simple which facilitate the task for the analysis business requirements. Firstly, the collection of requirements consists of diagnosis organization and requirement collection of defining the activities organization and contexts in order to collect the business requirements associated with each context for a specific activity. Then, the requirements specification requires their classification as strategic requirements, tactical and informational. Then, the treatment of these requirements is to establish associated models inter-goals according to the defined business context. Finally, filling models of informational goals formalization, explicit facts extraction and dimensions to establish the star schema. Secondly, we have presented a basic pattern that capitalizes models and activity diagrams of our approach. These patterns are the pattern product that contains the solution to establish or the pattern process that contains the solution process approach or both of combined.

In our future work, we will define a second set of patterns to ensure the adequate design for DIS. This basic pattern is intended to provide solutions and models approach which facilitates also the task of designers to design a DIS and explain the procedure to establish the DIS schema.